\documentclass[a4paper]{jpconf}
\usepackage{graphicx}
\usepackage{amsmath}
\usepackage{amssymb}

\begin{document}
\title{Exploring the Sun's core with BabyIAXO}

\author{Javier Galan in behalf of the IAXO Collaboration}

\address{Center for Astroparticles and High Energy Physics (CAPA), Universidad de Zaragoza, 50009 Zaragoza, Spain}

\ead{javier.galan@unizar.es}

\begin{abstract}
Axions are a natural consequence of the Peccei-Quinn mechanism, the most compelling solution to the strong-CP problem. Similar axion-like particles (ALPs) also appear in a number of possible extensions of the Standard Model, notably in string theories. Both, axions and ALPs, are very well motivated candidates for Dark Matter (DM), and they would be copiously produced at the sun’s core. A relevant effort during the last two decades has been the CAST experiment at CERN, the most sensitive axion helioscope to date. The International Axion Observatory (IAXO) is a large-scale 4th generation helioscope, and its primary physics goal is to extend further the search for solar axions or ALPs with a final signal to background ratio of about 5 orders of magnitude higher.

We briefly review here the astrophysical hints and models that will be at reach while searching for solar axions within the context of the IAXO helioscope search program, and in particular the physics under reach for BabyIAXO, an intermediate helioscope stage towards the full IAXO.
\end{abstract}

\section{Introduction}

The axion is a hypothetical particle that emerged as a natural solution to the strong CP problem\,\cite{CHENG19881}. In principle, the strong interaction, described by the quantum chromodynamics theory (QCD), does not forbid a CP violation, and the lagrangian of QCD allows a non-vanishing CP-violating term. Furthermore, since electroweak interactions violate CP, it is difficult to understand why nature has decided that strong interactions will not. The absence of CP-violation in strong interactions is quantified by the $\overline\theta$ parameter appearing at the corresponding CP-violating lagrangian term,

\begin{equation}
    \mathcal{L}_{\overline\theta} = -\overline\theta\frac{\alpha_s}{8\pi}G^a_{\mu\nu}\widetilde G^{a\mu\nu} 
\end{equation}

\noindent and it is measured experimentally to be unexpectedly close to zero, $|\overline\theta| < 1.8 \times 10^{-11}$, as it is deduced from the measurement of the electric dipole neutron moment,  $(0.0\pm1.1_{\mbox{stat}}\pm0.2_{\mbox{sys}})\times10^{-26}$\,e.cm., its corresponding upper limit $d_n < 1.8 \times 10^{-26}$\,e.cm.\,\cite{PhysRevLett.124.081803}, and a recent calculation 
in QCD, $|d_n|<10^{-15}\cdot\overline\theta$\,e.cm.\,\cite{PhysRevD.103.114507}. In principle, this phase could take any value between 0 and $2\pi$, since it is produced adding up QCD contributions of different nature, thus arising the question why this phase would have chosen a value such that the strong CP-violation term is canceled.

The Peccei-Quinn mechanism to restore CP conservation at the strong sector\,\cite{Peccei:1977hh,Peccei:1977ur} led soon into an interesting outcome - the axion - arising naturally as a dynamic solution to the fine tuning problem of the $\overline\theta$ parameter~\cite{Weinberg:1977ma,Wilczek:1977pj}, effectively, $\overline\theta$ is replaced by $\overline\theta + \left<A\right>/f_A$, where $\left<A\right>$ is the vacuum expectation value of the axion. 

The axion is a pseudo-scalar boson, and its properties are described by the theory through a unique parameter, the scale factor $f_a$, which must be much greater than the electroweak scale in order to circumvent the constrains imposed by accelerator based searches. Both, the couplings of axions to ordinary matter, and the axion mass, $m_a$, are inversely proportional to $f_a$. 

Two main theoretical models are considered as a reference for QCD axion searches, hadronic axions or KSVZ axions~\cite{KimModified,Shifman:1979if}, and DSFZ or GUT axions~\cite{Dine:1981rt,Zhitnitsky:1980tq}, which do not couple to hadrons at tree level. Furthermore, a plethora of axion-like particles (ALPs) emerge in different extensions to the SM. Their similarity with standard QCD axions makes that IAXO will be able to exploit its physics program to explore ALPs hinted regions other than the theoretical QCD models~\cite{Irastorza:2018dyq,Graham:2015ouw}.

\section{Physics motivation and astrophysical hints}

The parameter space of axions is strongly constrained by cosmological and astrophysical arguments, such as the evolution of stars or the expected cosmological axion abundance. As a consequence of its very weak couplings to ordinary matter the axion results to be a long-lived particle, and therefore, under certain conditions, it is considered to be a potential dark matter candidate. Particularly appealing are the regions of the parameter space that address the dark matter and QCD problem at once. 

The first studies including cosmological axion production through a mechanism called vacuum realignment mechanism\,\cite{PRESKILL1983127,ABBOTT1983133,DINE1983137} showed that the DM axion would overclose the universe for masses above $m_a\gtrsim10^{-6}$\,eV. This fact motivated the construction of axion haloscopes\,\cite{RevModPhys.75.777}, searching for axions with properties that match those of a model with a dominant axion contribution to dark matter. However, the axion mass that would lead to DM axions (those axions solving at the same time the dark matter puzzle) seems to be hard to predict by the theory. Its value has been relaxed in the latest years by additional production mechanisms, such as axion string radiation, predicting the mass of the DM axion could be high enough to be at the meV scale\,\cite{PhysRevD.82.123508,10.21468/SciPostPhys.10.2.050}. The next generation axion helioscopes will provide an apparatus able to probe those regions. In this context, it is also particularly attractive the search for the so-called ``ALP-miracle'', where an ALP at the meV region accounts for the dark matter of the universe and drives inflation at once\,\cite{ALPmiracle2017}. Scalar fields naturally appear as a way to explain a cosmological model with dark energy, such as quintessence fields\,\cite{PhysRevLett.75.2077}, chameleons\,\cite{PhysRevD.69.044026,PhysRevLett.93.171104} and other exotic candidates (see e.g.\,\cite{IRASTORZA201889} and references therein). The signatures produced by those exotic candidates share common features with the axion detection, and it has been already exploited in a campaign to detect chamaleons with the CERN Axion Solar Telescope (CAST)\,\cite{ANASTASSOPOULOS2015172}.

\begin{figure}[t]
	\begin{center}
		\includegraphics[width=0.75\textwidth]{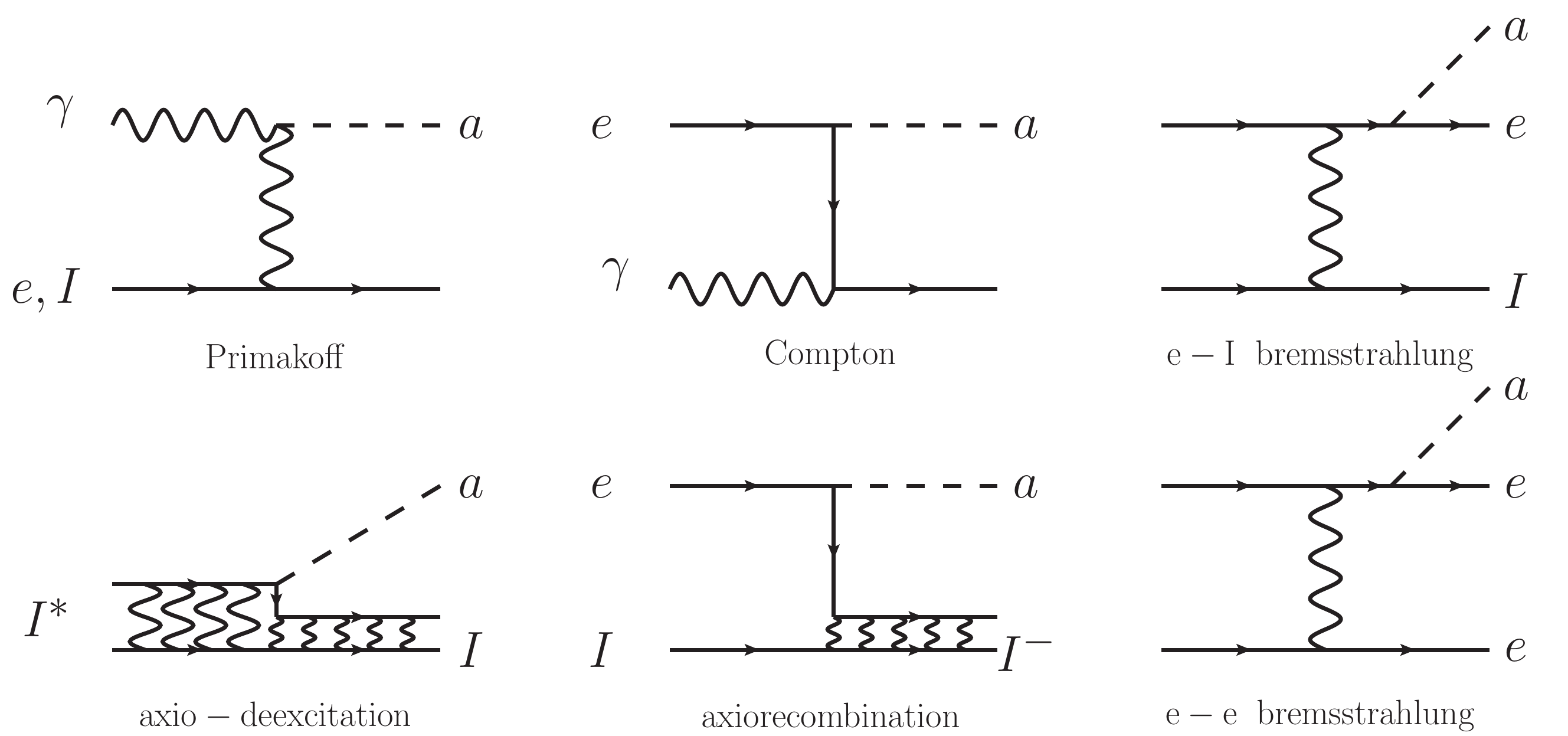} 
	\end{center}
	\caption{Feynman diagrams of the different processes responsible for axion production in the Sun. The Primakoff conversion of photons in the electromagnetic fields of the solar plasma is proportional to $g^2_{a\gamma}$, and it is present at almost any axion model. In non-hadronic models, such as DSFZ, non-negligible couplings to electrons open up new channels, such as Atomic axio-deexcitation and axio-recombination, axio-Bremsstrahlung in electron-ion or electron-electron collisions and Compton scattering with axion emission. Those additional channels are usually referred as the ABC solar axion flux, being the flux proportional to $g^2_{ae}$. Figure from ref~\cite{Redondo:2013wwa}.}
	\label{fig:processes}
\end{figure}

Observations from astrophysical origin produce even more stringent constrains\,\cite{PhysRevD.36.2211}. A star is a rich particle physics laboratory where we find many different physics processes and interactions where axions could be produced inside the stellar medium\,\cite{Redondo:2013wwa} (see Figure\,\ref{fig:processes}). The existence of the axion, or ALPs, could play a role on the evolution of the star since, just as neutrinos, they would escape the sun generating an additional energy loss channel. Depending on the axion coupling strength this additional energy loss channel might not be negligible, thus shortening the lifetime of the star. The age of our sun is well constrained by helioseismological observations and the measured neutrino flux, providing a first upper limit on the axion-photon coupling, $g_{a\gamma}\leq4.1\times10^{-10}$GeV$^{-1}$ (at 3$\sigma$)\,\cite{Vinyoles:2015aba}. Studying the evolution of stars through a Hertzsprung-Russel (HR) diagram we find even a more accurate measurement of the stars evolution by exploiting the statistics of stars at different evolutionary stages. The strongest bound comes from the Horizontal Branch (HB) to Red Giant Branch (RGB) ratio. The presence of an additional energy loss would reduce this ratio for a non-zero $g_{a\gamma}$ value. This fact leads to an improved upper limit of $g_{a\gamma}<0.66\times10^{-10}$GeV$^{-1}$ (at 2$\sigma$). Furthermore, the measured value for the HB to RGB ratio seems to be lower than expected. Considering that this effect could be coming from an energy loss through an axion Primakoff channel, we obtain what it is known as the ``HB hint'' value for $g_{a\gamma} = (0.29\pm0.18)\times10^{-10}$GeV$^{-1}$ (at 1$\sigma$)\,\cite{Ayala:2014pea}.

The Primakoff emission is dominant at Sun-like stars, where the density of the hot plasma is low enough to avoid production suppression due to the effective photon mass in the medium. Stars at different evolution stages, such as white dwarfs (WD) or neutron stars (NS) require other axion cooling channels, as axion-electron coupling, $g_{ae}$, or axion-nucleon coupling, $g_{aN}$. Another strong constrain arises for non-hadronic axion models through the evolution of WDs. The WD luminosity function (WDLF) does not reproduce the theoretical expectation, however, introducing an axion in the range  $g_{ae}=1.12-4.48\times10^{-13}$ corresponding to the axion mass range $m_a$cos$(\beta)=4-16$meV  (where cos$(\beta)$ is a free, model-dependent parameter that is usually set equal to unity) would reproduce the experimental results\,\cite{10.1093/mnras/sty1162}. This result is consistent with the cooling observed through an independent measurement of the change of period of a single WD, which leads to $g_{ae}=4.8\times10^{-13}$GeV$^{-1}$\,\cite{WDCorsico2012} for a $m_a$cos$(\beta)=17.1$meV. The cooling rate of few available single WDs measurements are also coherent with the presence of an anomalous cooling mechanism\,\cite{CorsicoReview}.

Another physics problem today is that the attenuation from extragalactic background light (EBL) is lower than expected, since high energy photons traveling long distances in the universe should interact with the EBL producing positron-electron pairs, thus introducing an effective cut-off on the energy gamma distribution observed on earth. The oscillation of those photons into ALPs and back again to photons in the presende of intergalactic magnetic fields could enlarge the effective optical depth and explain this anomaly\,\cite{MirizziTHint2017}. Some authors identify this effect -- that we denote as the universe transparency hint, ``T-hint'' -- with a low mass ALP of $m_a\sim7\times10^{-10}-5\times10^{-8}$\,eV, and an axion photon coupling of $g_{a\gamma}\sim 1.5\times10^{-11}-8.8\times10^{-10}$\,GeV$^{-1}$\,\cite{PhysRevD.96.051701} to explain the excess observed by the CIBER collaboration\,\cite{CIBER2017}, while other authors constrain the region nearby at the $g_{a\gamma}\sim10^{-12-11}$GeV$^{-1}$ level, for axion masses in the range $m_a\sim1-100$\,neV using HESS measurements\,\cite{BRUN201725} and Fermi-LAT observations\,\cite{CHENG2021136611}.


\section{Exploring the Sun interior with BabyIAXO}

\begin{figure}[h]
\begin{center}
\includegraphics[height=9cm]{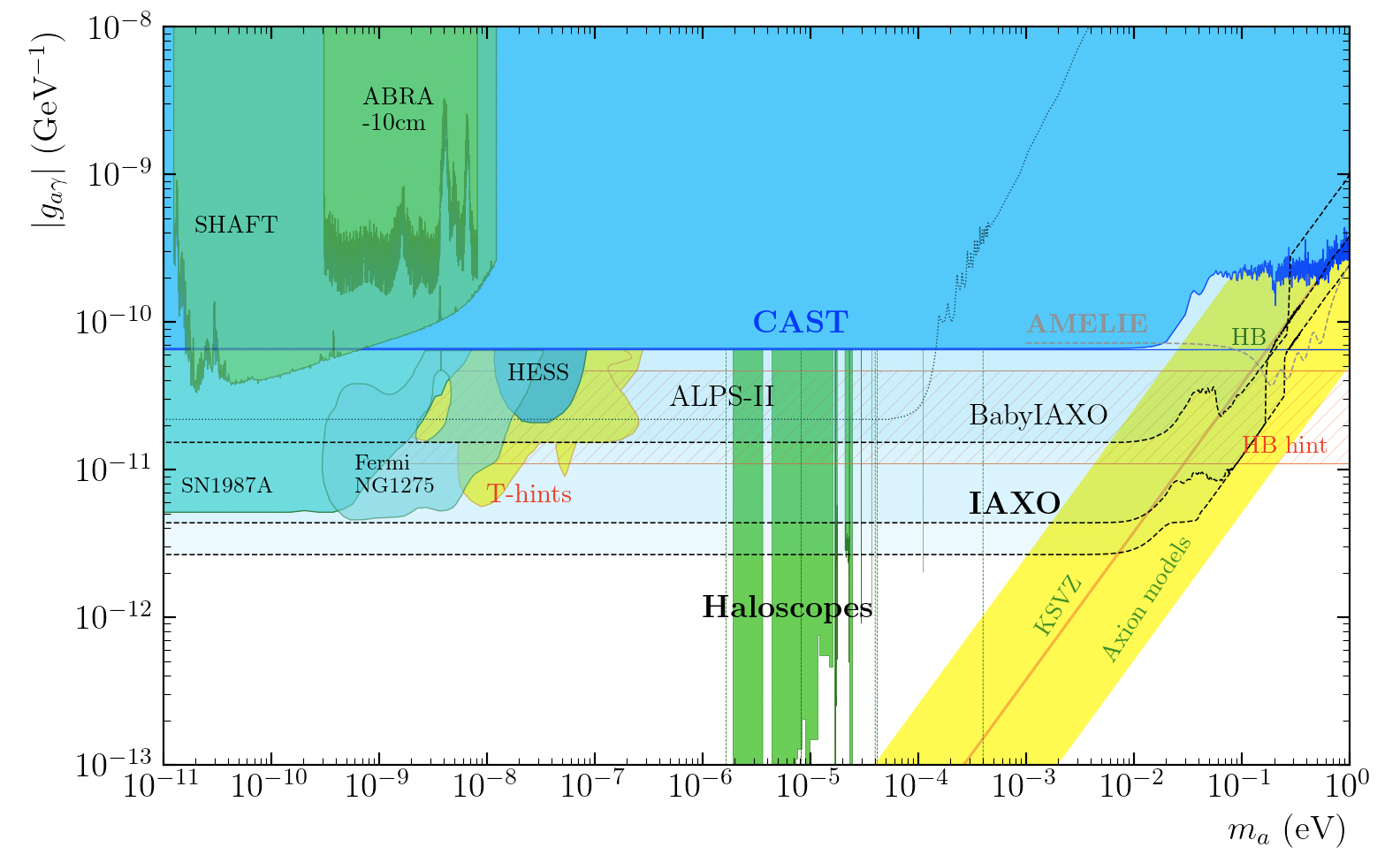}
\caption{\label{fig:exclusion} The prospects of the different IAXO setups and its sensitivity to $g_{a\gamma}$ as a function of the axion mass, $m_a$, including different hints and models described in the text, together with previous and planned experimental searches~\cite{Irastorza:2018dyq}.}
\end{center}
\end{figure}

IAXO is a 4th generation helioscope, pushing the sensitivity to solar axions to unprecedented levels, and surpassing those of the most sensitive axion helioscope up to date, the CERN Axion Solar Telescope (CAST)~\cite{Anastassopoulos:2017ftl}. The final IAXO configuration will allow to explore a new vast region of the axion parameter space, including a region predicted by QCD axions starting at the meV scale, and it will be sensitive enough to probe the mentioned astrophysical hints, as shown in Figure~\ref{fig:exclusion} together with the different IAXO stages towards the full IAXO\,\cite{Armengaud:2014gea,Armengaud_2019,BabyIAXOConceptual}.

Recent studies are increasing our understanding on the potential role of axions (or ALPs) in solar physics. In particular, calculations of axion to photon conversions in the presence of macroscopic magnetic fields demonstrate that a non-negligible contribution to the axion flux is expected from the different solar magnetic regions\,\cite{PhysRevD.101.123004,PhysRevD.102.123024}. This fact makes the next generation helioscopes an unprecedented tool to explore the interior of our closest star and to understand the mechanics of the internal magnetic field generation\,\cite{PhysRevD.102.043019}, in case the axion is discovered. Furthermore, using high energy resolution detectors to resolve axion atomic emissions from different isotopes would allow to better determine the solar heavy isotope composition, or metallicity\,\cite{PhysRevD.100.123020}.

\section{Conclusions}

Next generation axion helioscopes represent an opportunity to extend the Standard Model of particle physics. The discovery of the axion would entail a major breakthrough on the tools we have for the understanding of the universe, representing an additional component on multi-messenger astronomy. As briefly stated in this manuscript, the axion would be a potential candidate, together with neutrinos, to deeply study the physics of the Sun's interior.

If axions are discovered, the role of axion helioscopes will be crucial in the field. Most of the axion related processes allowed in the Sun interior could be differentiated with a dedicated setup at IAXO, allowing to independently measure axion couplings to photons, electrons, or nucleons\,\cite{PhysRevLett.75.3222}, leading to a complete description of the theory.

The IAXO physics program includes different phases towards the construction of the full IAXO helioscope potential. The first step towards the final stage is the construction of BabyIAXO that will serve as a proof of concept of the technologies chosen for its construction, and it will help to consolidate the different groups of expertise in the design and construction of the apparatus. BabyIAXO will allow to start probing and touch regions of the parameter space that are strongly motivated by observations, approaching the ``HB-hint'' and ``T-hint'', and covering partially the WD cooling anomaly hint. And at the same time it will cover regions of the parameter space at the meV scale, reaching regions favored by QCD axion models and/or connected with other fundamental physics problems, such as dark matter or inflation.




\section*{References}
\bibliography{iopart-num}

\providecommand{\newblock}{}
\begin{thebibliography}{10}
\expandafter\ifx\csname url\endcsname\relax
  \def\url#1{{\tt #1}}\fi
\expandafter\ifx\csname urlprefix\endcsname\relax\def\urlprefix{URL }\fi
\providecommand{\eprint}[2][]{\url{#2}}

\bibitem{CHENG19881}
Cheng H~Y 1988 {\em Physics Reports\/} {\bf 158} 1--89 ISSN 0370-1573

\bibitem{PhysRevLett.124.081803}
Abel C {\em et~al.\/} 2020 {\em Phys. Rev. Lett.\/} {\bf 124}(8) 081803

\bibitem{PhysRevD.103.114507}
Bhattacharya T, Cirigliano V, Gupta R, Mereghetti E and Yoon B 2021 {\em Phys.
  Rev. D\/} {\bf 103}(11) 114507

\bibitem{Peccei:1977hh}
Peccei R and Quinn H~R 1977 {\em Phys. Rev. Lett.\/} {\bf 38} 1440--1443

\bibitem{Peccei:1977ur}
Peccei R and Quinn H~R 1977 {\em Phys. Rev.\/} {\bf D16} 1791--1797

\bibitem{Weinberg:1977ma}
Weinberg S 1978 {\em Phys. Rev. Lett.\/} {\bf 40} 223--226

\bibitem{Wilczek:1977pj}
Wilczek F 1978 {\em Phys. Rev. Lett.\/} {\bf 40} 279--282

\bibitem{KimModified}
Kim J~E 1979 {\em Phys. Rev. Lett.\/} {\bf 43} 103

\bibitem{Shifman:1979if}
Shifman M~A, Vainshtein A~I and Zakharov V~I 1980 {\em Nucl. Phys.\/} {\bf
  B166} 493

\bibitem{Dine:1981rt}
Dine M, Fischler W and Srednicki M 1981 {\em Phys. Lett.\/} {\bf B104} 199

\bibitem{Zhitnitsky:1980tq}
Zhitnitsky A~R 1980 {\em Sov. J. Nucl. Phys.\/} {\bf 31} 260

\bibitem{Irastorza:2018dyq}
Irastorza I~G and Redondo J 2018  (\textit{Preprint} \eprint{1801.08127})

\bibitem{Graham:2015ouw}
Graham P~W, Irastorza I~G, Lamoreaux S~K, Lindner A and van Bibber K~A 2015
  {\em Ann. Rev. Nucl. Part. Sci.\/} {\bf 65} 485--514 (\textit{Preprint}
  \eprint{1602.00039})

\bibitem{PRESKILL1983127}
Preskill J, Wise M~B and Wilczek F 1983 {\em Physics Letters B\/} {\bf 120}
  127--132 ISSN 0370-2693

\bibitem{ABBOTT1983133}
Abbott L and Sikivie P 1983 {\em Physics Letters B\/} {\bf 120} 133--136 ISSN
  0370-2693

\bibitem{DINE1983137}
Dine M and Fischler W 1983 {\em Physics Letters B\/} {\bf 120} 137--141 ISSN
  0370-2693

\bibitem{RevModPhys.75.777}
Bradley R, Clarke J, Kinion D, Rosenberg L~J, van Bibber K, Matsuki S, M\"uck M
  and Sikivie P 2003 {\em Rev. Mod. Phys.\/} {\bf 75}(3) 777--817

\bibitem{PhysRevD.82.123508}
Wantz O and Shellard E~P~S 2010 {\em Phys. Rev. D\/} {\bf 82}(12) 123508

\bibitem{10.21468/SciPostPhys.10.2.050}
Gorghetto M, Hardy E and Villadoro G 2021 {\em SciPost Phys.\/} {\bf 10}(2) 50

\bibitem{ALPmiracle2017}
Daido R, Takahashi F and Yin W 2017  {\bf 2017} 044--044
  \urlprefix\url{https://doi.org/10.1088/1475-7516/2017/05/044}

\bibitem{PhysRevLett.75.2077}
Frieman J~A, Hill C~T, Stebbins A and Waga I 1995 {\em Phys. Rev. Lett.\/} {\bf
  75}(11) 2077--2080

\bibitem{PhysRevD.69.044026}
Khoury J and Weltman A 2004 {\em Phys. Rev. D\/} {\bf 69}(4) 044026

\bibitem{PhysRevLett.93.171104}
Khoury J and Weltman A 2004 {\em Phys. Rev. Lett.\/} {\bf 93}(17) 171104

\bibitem{IRASTORZA201889}
Irastorza I~G and Redondo J 2018 {\em Progress in Particle and Nuclear
  Physics\/} {\bf 102} 89--159 ISSN 0146-6410

\bibitem{ANASTASSOPOULOS2015172}
Anastassopoulos V {\em et~al.\/} 2015 {\em Physics Letters B\/} {\bf 749}
  172--180 ISSN 0370-2693

\bibitem{PhysRevD.36.2211}
Raffelt G~G and Dearborn D~S~P 1987 {\em Phys. Rev. D\/} {\bf 36}(8) 2211--2225

\bibitem{Redondo:2013wwa}
Redondo J 2013 {\em JCAP\/} {\bf 1312} 008 (\textit{Preprint}
  \eprint{1310.0823})

\bibitem{Vinyoles:2015aba}
Vinyoles N, Serenelli A, Villante F~L, Basu S, Redondo J and Isern J 2015 {\em
  JCAP\/} {\bf 1510} 015 (\textit{Preprint} \eprint{1501.01639})

\bibitem{Ayala:2014pea}
Ayala A, Dominguez I, Giannotti M, Mirizzi A and Straniero O 2014 {\em Phys.
  Rev. Lett.\/} {\bf 113} 191302 (\textit{Preprint} \eprint{1406.6053})

\bibitem{10.1093/mnras/sty1162}
Isern J, García-Berro E, Torres S, Cojocaru R and Catalán S 2018 {\em Monthly
  Notices of the Royal Astronomical Society\/} {\bf 478} 2569--2575 ISSN
  0035-8711

\bibitem{WDCorsico2012}
C{\'{o}}rsico A, Althaus L, Romero A, Mukadam A, Garc{\'{\i}}a-Berro E, Isern
  J, Kepler S and Corti M 2012  {\bf 2012} 010--010

\bibitem{CorsicoReview}
C{\'o}rsico A~H, Althaus L~G, Miller~Bertolami M~M and Kepler S~O 2019 {\em The
  Astronomy and Astrophysics Review\/} {\bf 27} 7

\bibitem{MirizziTHint2017}
Mirizzi A and Montanino D 2017 {\em JCAP\/} {\bf 2009} 004--004

\bibitem{PhysRevD.96.051701}
Kohri K and Kodama H 2017 {\em Phys. Rev. D\/} {\bf 96}(5) 051701

\bibitem{CIBER2017}
Matsuura S, Arai T, Bock J~J, Cooray A, Korngut P~M, Kim M~G, Lee H~M, Lee D~H,
  Levenson L~R, Matsumoto T, Onishi Y, Shirahata M, Tsumura K, Wada T and
  Zemcov M 2017 {\em The Astrophysical Journal\/} {\bf 839} 7

\bibitem{BRUN201725}
Brun P 2017 {\em Nuclear and Particle Physics Proceedings\/} {\bf 291-293}
  25--29 ISSN 2405-6014 “New eyes on the Universe” CRIS 2016 Cosmic Rays
  International Seminars Proceedings of the Cosmic Rays International Seminars

\bibitem{CHENG2021136611}
Cheng J~G, He Y~J, Liang Y~F, Lu R~J and Liang E~W 2021 {\em Physics Letters
  B\/} {\bf 821} 136611 ISSN 0370-2693

\bibitem{Anastassopoulos:2017ftl}
Anastassopoulos V {\em et~al.\/} (CAST) 2017 {\em Nature Phys.\/} {\bf 13}
  584--590 (\textit{Preprint} \eprint{1705.02290})

\bibitem{Armengaud:2014gea}
Armengaud E, Avignone F, Betz M, Brax P, Brun P {\em et~al.\/} 2014 {\em
  JINST\/} {\bf 9} T05002 (\textit{Preprint} \eprint{1401.3233})

\bibitem{Armengaud_2019}
Armengaud E {\em et~al.\/} 2019 {\em Journal of Cosmology and Astroparticle
  Physics\/} {\bf 2019} 047--047

\bibitem{BabyIAXOConceptual}
Abeln A {\em et~al.\/} 2021 {\em Journal of High Energy Physics\/} {\bf 2021}
  137

\bibitem{PhysRevD.101.123004}
Caputo A, Millar A~J and Vitagliano E 2020 {\em Phys. Rev. D\/} {\bf 101}(12)
  123004

\bibitem{PhysRevD.102.123024}
Guarini E, Carenza P, Gal\'an J, Giannotti M and Mirizzi A 2020 {\em Phys. Rev.
  D\/} {\bf 102}(12) 123024

\bibitem{PhysRevD.102.043019}
O'Hare C~A~J, Caputo A, Millar A~J and Vitagliano E 2020 {\em Phys. Rev. D\/}
  {\bf 102}(4) 043019

\bibitem{PhysRevD.100.123020}
Jaeckel J and Thormaehlen L~J 2019 {\em Phys. Rev. D\/} {\bf 100}(12) 123020

\bibitem{PhysRevLett.75.3222}
Moriyama S 1995 {\em Phys. Rev. Lett.\/} {\bf 75}(18) 3222--3225

\end{thebibliography}

\end{document}